\documentclass[titlepage]{article}
\usepackage{amsmath}
\usepackage{amsfonts}
\usepackage{amssymb}

\begin{document}

\title{Static  Anisotropic Solutions to Einstein Equations  with  a \\
Nonlocal Equation of State}
\author{\textbf{H. Hern\'{a}ndez}\\{\textit{Laboratorio de F\'{i}sica Te\'{o}rica, Departamento de F\'{i}sica,} }\\{\textit{Facultad de Ciencias, Universidad de Los Andes,}}\\{\textit{M\'{e}rida 5101, Venezuela} }\\and \\\textit{Centro Nacional de C\'alculo Cient\'{\i}fico}\\\textit{Universidad de Los Andes (CeCalCULA),}\\\textit{Corporaci\'{o}n Parque Tecnol\'{o}gico de M\'{e}rida, }\\\textit{M\'{e}rida 5101, Venezuela}
\and \textbf{L.A. N\'{u}\~{n}ez}\\{\textit{Centro de Astrof\'{i}sica Te\'{o}rica,} \textit{Facultad de Ciencias,
}}\\{\textit{Universidad de Los Andes,} \textit{M\'{e}rida 5101, Venezuela} }\\and \\\textit{Centro Nacional de C\'alculo Cient\'{\i}fico}\\\textit{Universidad de Los Andes (CeCalCULA),}\\\textit{Corporaci\'{o}n Parque Tecnol\'{o}gico de M\'{e}rida, }\\\textit{M\'{e}rida 5101, Venezuela}}
\date{November 2000}
\maketitle
\begin{abstract}
We present a general method to obtain static anisotropic spherically symmetric
solutions, satisfying a nonlocal equation of state, from known density
profiles. This equation of state describes, at a given point, the components
of the corresponding energy-momentum tensor not only as a function at that
point, but as a functional throughout the enclosed configuration. In order to
establish the physical aceptability of the proposed static family of solutions
satisfying nonlocal equation of state,\textit{ }we study the consequences
imposed by the junction and energy conditions for anisotropic fluids in
bounded matter distribution. It is shown that a general relativistic
spherically symmetric bounded distributions of matter, at least for certain
regions, could satisfy a nonlocal equation of state.
\end{abstract}

\section{Introduction}

The structure of a relativistic star is believed to be rather complicated
(solid crust, superfluid interior, different exotic phases transitions).
Despite that its bulk properties seem to be computed with reasonable accuracy
making several simplifying assumptions, the true equation of state that
describes the properties of matter at densities higher than nuclear
($\approx10^{14}$ $gr/cm.^{3}$) is essentially unknown. This is mainly due to
our inability to verify experimentally the theories that should describe the
microphysics of nuclear matter at such high densities
\cite{Demianski1985,ShapiroTeukolsky1983,KippenhahnWeigert1990,Glendening2000}%
. Presently, what is known comes from the experimental insight and
extrapolations from the ultra high energy accelerators and experimental cosmic
physics (see \cite{Glendening2000,ArnettBowers1977,SalgadoEtal1994} and
references therein). Having this uncertainty in mind, it seems be reasonable
to explore what is allowed by the laws of physics, in particular, within the
framework of the theory of General Relativity and considering spherical and
axial symmetries.

Classical continuum theories are based on the assumption that the state of a
body is determined entirely by the behavior of an arbitrary infinitesimal
neighborhood centered at any of its material points. Furthermore, there is
also a premise that any small piece of the material can serve as a
representative of the entire body in its behavior and, hence the governing
balance laws are assumed to be valid for every part of the body, no matter how
small. Clearly, the influence of the neighborhood on motions of the material
points, emerging as a result of the interatomic interaction of the rest of the
body, is neglected. Moreover, the isolation of an arbitrary small part of the
body to represent the entire one clearly ignores the effects of the action of
the applied load at distance. These applied loads are important because their
transmissions from one part of the body to another, through their common
boundaries, affect the motions and hence the state of the body at every point.

The relevance of long-range or nonlocal outcomes on the mechanical properties
of materials are well known. The main ideas of non local a continuum were
introduced during the 1960s and are based on considering the stress to be a
function of the mean of the strain from a certain representative volume of the
material centered at that point. From that time there have been many
situations of common occurrence wherein nonlocal effects seem to dominate the
macroscopic behavior of matter. Interesting problems coming from a wide
variety of areas such as damage and cracking analysis of materials, surface
phenomena between two liquids or two phases, mechanics of liquid crystals,
blood flow, dynamics of colloidal suspensions seem to demand this type of
nonlocal approach which has made this area very active concerning recent
developments in material and fluid science and engineering (see
\cite{Narasimhan1993} and references therein).

In a recent work \cite{HernandezNunezPercoco1998}, it is shown that under
particular circumstances a general relativistic spherically symmetric
anisotropic distribution of matter could satisfy a \textit{Nonlocal Equation
of State} (\textit{NLES} from now on). Some types of dynamic bounded matter
configurations having a \textit{NLES}, with constant gravitational potentials
at the surface, admit a Conformal Killing Vector and fulfill the energy
conditions for anisotropic imperfect fluids. We also developed several
analytical and numerical models for collapsing radiating anisotropic spheres
in general relativity.

The present paper is focussed in a more general family of \textit{NLES} and in
determining the conditions under it could represent reasonable bounded matter
distribution in General Relativity.

The static limit of the particular \textit{NLES} considered in reference
\cite{HernandezNunezPercoco1998} can be written as%

\begin{equation}
P_{r}(r)=\rho(r)-\frac{2}{r^{3}}\int_{0}^{r}\bar{r}^{2}\rho(\bar
{r})\ \mathrm{d}\bar{r}\ +\frac{\mathcal{C}}{2\pi r^{3}}\ ; \label{globalst}%
\end{equation}
where $\mathcal{C}$ is an arbitrary integration constant. It is clear in
equation (\ref{globalst}) a collective behavior on the physical variables
$\rho(r)$ and $P_{r}(r)$ is present. The pressure $P_{r}(r)$ is not only a
function of the energy density, $\rho(r),$ at that point but also its
functional throughout the rest of the configuration. Any change in the
pressure takes into account the effects of the variations of the energy
density within an entire volume.

An additional physical insight of the meaning of the nonlocality for this
particular equation of state can be gained by considering equation
(\ref{globalst}) re-written as
\begin{equation}
P_{r}(r)=\rho(r)-\frac{2}{3}\left\langle \rho\right\rangle _{r}+\frac
{\mathcal{C}}{r^{3}}\ ,\qquad\mathrm{with\quad}\left\langle \rho\right\rangle
_{r}=\frac{\int_{0}^{r}4\pi\bar{r}^{2}\rho(\bar{r})\ \mathrm{d}\bar{r}}
{\frac{4\pi}{3}r^{3}}\ =\frac{M(r)}{V(r)}\ \label{averagedens1}%
\end{equation}
Clearly the nonlocal term represents an average over the function $\rho(r)$
within the volume enclosed by the radius $r$. Moreover, equation
(\ref{averagedens1}) can be easily rearranged as
\begin{equation}
P(r)=\frac{1}{3}\rho(r)+\frac{2}{3}\ \left(  \rho(r)-\left\langle
\rho(r)\right\rangle \right)  \ +\frac{\mathcal{C}}{r^{3}}=\frac{1}{3}
\rho(r)+\frac{2}{3}\ \mathbf{\sigma}_{\rho}+\frac{\mathcal{C}}{r^{3}}\ ,
\label{sigmatmunu}%
\end{equation}
where we have used the concept of statistical standard deviation
$\mathbf{\sigma}_{\rho}$ from the local value of energy density. Furthermore,
we may write:%

\begin{equation}
P_{r}(r)=\mathcal{P}(r)+2\mathbf{\sigma}_{\mathcal{P}(r)}+\frac{\mathcal{C}
}{r^{3}}\ \quad\mathrm{where\quad}\left\{
\begin{array}
[c]{l}%
\mathcal{P}(r)=\frac{1}{3}\rho(r)\\
\\
\mathbf{\sigma}_{\mathcal{P}(r)}=\left(  \frac{1}{3}\rho(r)-\frac{1}%
{3}\left\langle \rho\right\rangle _{r}\right)  =\left(  \mathcal{P}%
(r)-{\bar{\mathcal{P}}}(r)\right)
\end{array}
\right.  \label{sigma0}%
\end{equation}
Therefore, if at a particular point within the distribution the value of the
density, $\rho(r),$ gets very close to its average $\left\langle
\rho(r)\right\rangle $ the equation of state of the material becomes similar
to the typical radiation dominated environment, $P_{r}(r)\approx
\mathcal{P}(r)\equiv\frac{1}{3}\rho(r).$

The structure of this contribution is the following: first we give the general
conventions and the field equations; secondly in Section III the nonlocal
equation of state for anisotropic static models is presented; following is
Section IV where we consider the consequences imposed by the junction and
energy conditions; and finally the new solutions are shown in the Section V.

\section{The Einstein Field Equations}

To explore the feasibility of nonlocal equations of state for bounded
configurations in General Relativity, we shall consider a static spherically
symmetric anisotropic distribution of matter with an energy-momentum
represented by $\mathbf{T}_{\nu}^{\mu}~=~{diag}\,~(\rho,-P_{r},-P_{\perp
},-P_{\perp})$. Here $\rho$ is the energy density, $P_{r}$ the radial pressure
and $P_{\perp}$ the tangential pressure. Although the perfect pascalian fluid
assumption (i.e. $P_{r}~=~P_{\perp}$) is supported by solid observational and
theoretical grounds, an increasing amount of theoretical evidence strongly
suggests that, for certain density ranges, a variety of very interesting
physical phenomena may take place giving rise to local anisotropy (see
\cite{HerreraSantos1997} and references therein).

We adopt standard Schwarzschild coordinates $(t,r,\theta,\phi)$ where the line
element can be written as
\begin{equation}
\mathrm{d}s^{2}=e^{2\nu(r)}\mathrm{d}t^{2}-e^{2\lambda(r)}\mathrm{d}%
r^{2}-r^{2}\mathrm{d}\Omega^{2}\,, \label{metrica}%
\end{equation}
with $d\Omega^{2}\equiv d\theta^{2}+\sin^{2}\theta d\phi^{2}$ , the solid angle.

The resulting Einstein equations are:
\begin{align}
8\pi\rho &  =\frac{1}{r^{2}}+\frac{e^{-2\lambda}}{r}\left[  2\lambda^{\prime
}-\frac{1}{r}\right]  \,,\label{ee1}\\
-8\pi P_{r}  &  =\frac{1}{r^{2}}-\frac{e^{-2\lambda}}{r}\left[  2\nu^{\prime
}+\frac{1}{r}\right]  \,\quad\qquad\mathrm{and}\label{ee2}\\
-8\pi P_{\perp}  &  =e^{-2\lambda}\left[  \frac{\lambda^{\prime}}{r}-\frac
{\nu^{\prime}}{r}-\nu^{\prime\prime}+\nu^{\prime}\lambda^{\prime}-(\nu
^{\prime})^{2}\right]  \,, \label{ee3}%
\end{align}
where primes denote differentiation with respect to $r$.

Using equations (\ref{ee2}) and (\ref{ee3}), or equivalently the conservation
law ${\mathbf{T}_{\nu}^{\mu}}_{;\mu}~=~0$, we obtain the hydrostatic
equilibrium equation for anisotropic fluids
\begin{equation}
P_{r}^{\prime}=-\left(  \rho+P_{r}\right)  \nu^{\prime}+\frac{2}{r}\left(
P_{\perp}-P_{r}\right)  \,. \label{eeh}%
\end{equation}
It can be formally integrated to give
\begin{equation}
e^{-2\lambda}=1-2\frac{m(r)}{r}\,, \label{explam}%
\end{equation}
where the mass function $m(r)$ defined by
\begin{equation}
m(r)=4\pi\,\int_{0}^{r}\rho\,{{\bar{r}}^{2}\,\mathrm{d}{\bar{r}}}\,,
\label{eme}%
\end{equation}
is the mass inside a sphere of radius $r$ as seen by a distant observer.
Finally, from(\ref{eeh}), (\ref{ee1}) and (\ref{ee2}) the anisotropic
Tolman-Oppenheimer-Volkov (TOV) equation \cite{BowerLiang1974} can be written
as
\begin{equation}
\frac{\mathrm{d\,}P_{r}}{\mathrm{d}\,r}=-\left(  \rho+P_{r}\right)  \left(
\frac{m+4\pi r^{3}P_{r}}{r\left(  r-2m\right)  }\right)  +\frac{2}{r}\left(
P_{\perp}-P_{r}\right)  \,. \label{anitov}%
\end{equation}
Obviously, in the isotropic case $(P_{\perp}=P_{r})$ it becomes the usual TOV equation.

\section{A Family of Solutions with a \textit{NLES}}

In this section we are going to present a family of static solution of the
Einstein Equation satisfying a \textit{NLES} Defining the new variables:
\begin{equation}
e^{2\nu\left(  r\right)  }=h\left(  r\right)  \,e^{4\beta\left(  r\right)
},\quad\mathrm{and}\qquad e^{2\lambda\left(  r\right)  }=\frac{1}{h\left(
r\right)  };\quad\mathrm{\qquad with}\qquad h(r)\equiv1-2\frac{m\left(
r\right)  }{r}\ ,\label{elemetric}
\end{equation}
the above metric (\ref{metrica}) can be re-written as

\begin{equation}
\mathrm{d}s^{2}=h\left(  r\right)  \,e^{4\beta(r)}\mathrm{d}t^{2}-\frac
{1}{h(r)}dr^{2}-r^{2}\mathrm{d}\Omega^{2}\,, \label{metrica2}
\end{equation}
and the resulting Einstein Equations are:
\begin{align}
8\pi\rho &  =\frac{1-h-h^{\prime}r}{r^{2}}\,,\label{eeg1}\\
8\pi P_{r}  &  =-\frac{1-h-h^{\prime}r}{r^{2}}+\frac{4\,h\,\beta^{\prime}}
{r}\quad\qquad\mathrm{and}\label{eeg2}\\
8\pi P_{\perp}  &  =\frac{\,h^{\prime}+2\,h\,\beta^{\prime}}{r}+\frac{1}
{2}\left[  \ h^{\prime\prime}+4\,h\,\beta^{\prime\prime}+6\,h^{\prime}
\beta^{\prime}+8\,h\,\left(  \beta^{\prime}\right)  ^{2}\right]  \,.
\label{eeg3}%
\end{align}
Now, if equation (\ref{globalst}) is re-stated as
\begin{equation}
\rho-3P_{r}+r\left(  \rho^{\prime}-P_{r}^{\prime}\right)  =0\ , \label{eeg}%
\end{equation}
we have, from (\ref{eeg}) and by using (\ref{eeg1}) - (\ref{eeg2}),
\begin{equation}
\frac{2}{r}\left(  h^{\prime}+2h\beta^{\prime}\right)  +h^{\prime\prime
}+2\beta^{\prime}h^{\prime}+2h\beta^{\prime\prime}=0\ .
\end{equation}
It can be formally integrated yielding
\begin{equation}
\beta(r)=\frac{1}{2}\ln\left(  \frac{\mathrm{C}}{h}\right)  + \int
\frac{\mathcal{C}}{r^{2}\,h}\ \mathrm{d}\,r+C_{1} \label{beta}%
\end{equation}
where $\mathrm{C}$ and $C_{1}$ are arbitrary integration constants.

The corresponding Einstein equations\ in terms of the metric elements
(\ref{elemetric}) are:
\begin{align}
8\pi\rho &  =\frac{2m^{\prime}}{r^{2}},\label{rho}\\
8\pi P  &  =8\pi\rho-\frac{4\left(  m-\mathcal{C}\right)  }{r^{3}},\quad
\qquad\mathrm{and}\label{Prad}\\
8\pi P_{\perp}  &  =\frac{m^{\prime\prime}r^{2}\left(  r-2m\right)
+2r^{2}m^{\prime}\left(  m^{\prime}-1\right)  -2m\left(  m-r\right)  -2
{\mathcal{C}}\left(  r+2m-3rm^{\prime}\right)  +4{\mathcal{C}}^{2}}%
{r^{3}\left(  r-2m\right)  } \label{Ptang}%
\end{align}

At this point, equation (\ref{beta}) deservers several comments.

\begin{itemize}
\item  Firstly, if we set ${\mathcal{C}}~=~C_{1}~=~0$ , a particular solution
is found, i.e.,
\begin{equation}
\beta(r)=\frac{1}{2}\ln\left(  \frac{\mathrm{C}}{h}\right)  ,
\end{equation}
with $\mathrm{C}$ a constant parameter. We have considered this family of
solutions in a previous work \cite{HernandezNunezPercoco1998} and some of
their geometric properties and several collapsing models of radiating
anisotropic spheres were presented.

\item  The second comment concerning equation (\ref{beta}) is the approach we
have followed in order to obtain static anisotropic solutions having a
\textit{NLES}. It is clear that if the profile of the energy density,
$\rho(r)$, is provided, the metric elements $h(r)$ and $\beta(r)$ can be
calculated through (\ref{eme}), (\ref{elemetric}) and (\ref{beta}). Therefore
we can device a consistent method to obtain static solutions having
\textit{NLES} from a known static solutions.
\end{itemize}

It is clear that the metric elements\ describing bounded matter distribution
should fulfill the junctions conditions and the physical variables coming from
the energy momentum tensor are only restricted by some elementary criteria of
physical acceptability and the hydrostatic equilibrium equation (\ref{anitov}%
). The next section is devoted to list those conditions for anisotropic fluids.

\section{Energy and Junctions Conditions}

Most of the spherically symmetric exact solutions\ found in the literature do
not represent physically ``realistic'' fluids (see for examples, two
interesting and complementaries reviews on this subject \cite{FinchSkea} and
\cite{DelgatyLake1998}).\ The condition of what is a realistic fluid is
subjective and depends from author to author on which of the energy conditions
are considered valid. In order to establish the physical aceptability of the
proposed static family of solutions (\ref{beta}) satisfying a \textit{NLES,
(}\ref{globalst}) we shall study consequences imposed by the junction and
energy conditions for anisotropic fluids on bounded matter distribution.

In order to be matched to the exterior Schwarzschild solution,
\begin{equation}
\mathrm{d}s^{2}=\left(  1-2\frac{M}{r}\right)  \mathrm{d}t^{2}-\left(
1-2\frac{M}{r}\right)  ^{-1}\mathrm{d}r^{2}-r^{2}\mathrm{d}\Omega^{2}\,,
\end{equation}
the interior metric (\ref{metrica2}) should satisfy the following conditions
at the surface of the sphere $r=a:$
\begin{equation}
\beta\left(  a\right)  =\beta_{a}=0,\qquad\qquad m(a)=M\qquad\mathrm{and}%
\qquad P_{r}(a)=0 \label{cda}%
\end{equation}
thus equation (\ref{Prad}) leads to
\begin{equation}
{\mathcal{C}}=M-2\pi a^{3}\rho\left(  a\right)
\end{equation}

The elementary criteria we use to characterize a physically meaningful
anisotropic fluid are:

\begin{enumerate}
\item  The pressure must be positive:
\begin{equation}
P_{r}\geq0\qquad\mathrm{and}\qquad P_{\perp}\geq0 \label{presposit}%
\end{equation}

\item  The pressure gradient must be negative:
\begin{equation}
\frac{\partial P_{r}}{\partial r}\leq0 \label{Dpres}%
\end{equation}

\item  The speed of sound should be subluminal:
\begin{equation}
v_{s}\equiv\frac{\partial P_{r}}{\partial\rho}\leq1 \label{velsonid}%
\end{equation}

\item  The trace of the energy-momentum tensor must be positive
(\textit{Strong Energy Condition}):
\begin{equation}
\rho>P_{r}+2\,P_{\perp} \label{strgener}%
\end{equation}

\item  The density must be larger that the pressure (\textit{Dominant Energy
Condition}):
\begin{equation}
\rho>P_{r}\qquad\mathrm{and}\qquad\rho>P_{\perp} \label{domener}%
\end{equation}
\end{enumerate}

\section{The Method and \textit{NLES} Static Solutions}

In the present section we state a general method to obtain \textit{NLES}
static anisotropic spherically symmetric solution from density profiles.

\begin{enumerate}
\item  Select a static density profile $\rho(r)$, from a known static
solution. Thus, the mass function can be obtained through equation (\ref{eme})
and by the junction condition (\ref{cda}) fulfills the continuity of $h(a)$
giving the expression for the total mass $m(a)=M$.

\item  Following, the metric coefficient $\beta(r)$ can be found by using
equation (\ref{beta}).

\item  Finally, Einstein equations (\ref{Prad}) through (\ref{Ptang}) provide
the expressions for the radial and tangential pressures, $P_{r}$ and
$P_{\perp}$, respectively. The integration constants are obtained as
consequences of the junction conditions at the boundary surface, $r=a$, i.e.
$\beta(a)=0,$ $m(a)=M$ and $P_{r}(a)=0.$

\item  Once we have a candidate for a bounded matter distribution having a
\textit{NLES} we should explore where, within the configuration, all the above
conditions (equations (\ref{presposit}) through (\ref{domener})) are valid.
\end{enumerate}

In order to illustrate the above procedure we shall work out seven static
solutions: \textit{Tolman VI}, and \textit{Tolman-V} \cite{Tolman1939}, an
static solution proposed by B. W. Stewart \cite{Stewart1982}, three solutions
proposed by M. C. Durgapal \cite{Durgapal1982} and the M. K. Gokhroo and A. L.
Mehra solution \cite{GokhrooMehra1994}, (see Appendix). Tolman and Durgapal
solutions are borrowed from isotropic static distribution of matter but
Stewart and Gokhroo \& Mehra solutions describe static bounded anisotropic
ones. Stewart solution outlines a conformally flat anisotropic distribution
which is more stable than its isotropic counterpart.

The Gokhroo and Mehra solution corresponds to an anisotropic fluid with
variable density and under some circumstances \cite{Martinez1996}, represents
densities and pressures given rise to an equation of state similar to the
Bethe-B\"{o}rner-Sato newtonian equation for nuclear matter
\cite{Demianski1985,ShapiroTeukolsky1983}.

The expressions for the starting energy density can be summarized as follows:

Singular models:

\begin{center}%
\begin{tabular}
[c]{|c|c|}\hline
\textbf{Equation of State} & \textbf{Density $\rho$}\\\hline
\textbf{Tolman V} & $\frac{1}{8 \pi\,}\, \left[  \frac{3}{7 r^{2}}+\frac{10}{3
R^{2}}\, \left(  \frac{r}{R}\right)  ^{\frac{1}{3}}\right]  $\\\hline
\textbf{Tolman VI} & $\frac{3 \sigma}{56\,\pi\,r^{2}} $\\\hline
\textbf{Stewart } & $\frac{1}{8\,\pi r^{2}}\frac{\left(  e^{2Kr}-1\right)
\left(  e^{4Kr}+8Kre^{2Kr}-1\right)  }{\left(  e^{2Kr}+1\right)  ^{3}}%
$\\\hline
\end{tabular}
\end{center}

No singular models:

\begin{center}%
\begin{tabular}
[c]{|c|c|}\hline
\textbf{Equation of State} & \textbf{Density $\rho$}\\\hline
\textbf{Durgapal} (n=1) & $\frac{C}{8\,\pi} \left[  \frac{1-3K-3Kx}{\left(
1+2\,x\right)  } + \frac{2\left(  1+Kx \right)  }{\left(  1+2\,x\right)  ^{2}%
}\right]  $\\\hline
\textbf{Durgapal} (n=2) & $-\frac{C}{8\,\pi}\frac{K\left(  3+5\,x\right)
}{\left(  1+3\,x\right)  ^{\frac{5}{3}}}$\\\hline
\textbf{Durgapal} (n=3) & $\frac{C}{8\,\pi\left(  1+\,x\right)  ^{2}}\left[
\frac{3}{2}\left(  3+x\right)  -\frac{3K\left(  1+3\,x\right)  }{\left(
1+4\,x\right)  ^{\frac{3}{2}}}\right]  $\\\hline
\textbf{Gokhroo \& Mehra} & $\sigma\left[  1-K\frac{r^{2} }{a^{2}}\right]
$\\\hline
\end{tabular}
\end{center}

The parameters that characterize bounded configurations: mass $M$, $M/a$,
boundary redshift $z_{a}$, surface density $\rho_{a}$ and central density
$\rho_{c}$ the are the shown in the following table. Let us consider stars
with radius of $10$ Km.

\begin{center}%
\begin{tabular}
[c]{|c|c|c|c|c|}\hline
\textbf{Equation of State} & M/a & $M$ $(M_{\odot})$ & $z_{a}$ & $\rho_{a}$
$\times$ $10^{14}\,(gr.cm^{-3})\,$\\\hline
\textbf{TolmanV} & 0.25 & 1.69 & 0.4 & 3.57\\\hline
\textbf{TolmanVI} & 0.25 & 1.70 & 0.4 & 2.68\\\hline
\textbf{Stewart} & 0.32 & 2.15 & 0.6 & 6.80\\\hline
\end{tabular}
\end{center}

and for the models with central density:

\begin{center}%
\begin{tabular}
[c]{|c|c|c|c|c|c|}\hline
\textbf{Equation of State} & M/a & $M$ $(M_{\odot})$ & $z_{a}$ & $\rho_{a}$
$\times$ $10^{14}\,(gr.cm^{-3})$ & $\rho_{c}$ $\,\times10^{15}$ $(gr.cm^{-3}%
)$\\\hline
\textbf{Durgapal }(n=1) & 0.25 & 1.69 & 0.4 & 5.36 & 2.00\\\hline
\textbf{Durgapal }(n=2) & 0.33 & 2.26 & 0.7 & 7.15 & 2.70\\\hline
\textbf{Durgapal }(n=3) & 0.38 & 2.54 & 1.0 & 8.04 & 2.41\\\hline
\textbf{Gokhroo \& Mehra} & 0.20 & 1.35 & 0.3 & 4.29 & 0.96\\\hline
\end{tabular}
\end{center}

Within the above configurations the range of validity for the \textit{NLES} are

\begin{center}%
\begin{tabular}
[c]{|c|c|}\hline
\textbf{Equation of State} & $r\;(Km)$\\\hline
\textbf{Tolman V} & $7.8\leq r\leq10.0$\\\hline
\textbf{Tolman VI} & $7.5\leq r\leq10.0$\\\hline
\textbf{Stewart} & $0.7\leq r\leq10.0$\\\hline
\textbf{Durgapal}(n=1) & $0.0\leq r\leq10.0$\\\hline
\textbf{Durgapal}(n=2) & $0.0\leq r\leq10.0$\\\hline
\textbf{Durgapal}(n=3) & $0.0\leq r\leq10.0$\\\hline
\textbf{Gokhroo \& Mehra} & $1.1\leq r\leq8.2$\\\hline
\end{tabular}
\end{center}

\section{Acknowledgments}

We are indebted to J. Fl\'{o}rez L\'{o}pez for pointing out us the relevance
of nonlocal theories in modern classical continuum mechanics. We also
gratefully acknowledge the financial support of the Consejo de Desarrollo
Cient\'{i}fico Human\'{i}stico y Tecnol\'{o}gico de la Universidad de Los Andes.\newpage

\section{\textbf{Appendix}}

\subsection{Tolman V-like solution}

We shall now take the solution V of Tolman as providing the first of the
examples of a sphere of fluid surrounded by empty space. The density is given by:%

\begin{equation}
\rho=\frac{1}{8 \pi\,}\,\left[  \frac{3}{7 r^{2}}+\frac{10}{3 R^{2}} \,\left(
\frac{r}{R}\right)  ^{\frac{1}{3}}\right]
\end{equation}

>From equations (\ref{eme}) and\ (\ref{globalst}) we obtain the expressions
for the mass and radial pressure,%

\begin{equation}
m=\frac{r^{3}}{2}\left[  \frac{3}{7r^{2}}+\frac{1}{R^{2}}\left(  \frac{r}
{R}\right)  ^{\frac{1}{3}}\right]  \qquad\mathrm{and}\qquad P_{r}=\rho
-\frac{1}{4\pi}\left[  \frac{3}{7r^{2}}+\frac{1}{R^{2}}\left(  \frac{r}
{R}\right)  ^{\frac{1}{3}}\right]  +\frac{\mathcal{C}}{2\pi r^{3}}%
\end{equation}
respectively.

>From the boundary conditions, $C_{1}$ and $R$ are found to be
\begin{equation}
{\mathcal{C}}=\frac{a}{84}\left[  9-28\left(  \frac{a}{R}\right)  ^{\frac
{7}{3}}\right]  \qquad\mathrm{and}\qquad R=\frac{7^{\frac{3}{7}}a^{\frac
{10}{7}}}{\left(  14M-3a\right)  ^{\frac{3}{7}}}%
\end{equation}

The following table contains the ranks of validity for the different criteria
indicated in the section IV

\begin{center}%
\begin{tabular}
[c]{|c|c|c|c|c|c|c|c|c|}\hline\hline
& $\rho\geq0$ & $P_{r}\geq0$ & $P_{\perp}\geq0$ & $\frac{\partial P_{r}%
}{\partial r}<0$ & $\frac{\partial P_{r}}{\partial\rho}\leq1$ & $\rho
>P_{r}+2\,P_{\perp}$ & $\rho>P_{r}$ & $\rho>P_{\perp}$\\\hline
$r\in$ & $\left(  0,\ 10.0\right]  $ & $\left(  0,\ 10.0\right]  $ & $\left.
\begin{array}
[c]{c}%
\left(  0,\ 0.3\right] \\
\cup\\
\left[  7.8,\ 10.0\right]
\end{array}
\right.  $ & $\left(  0,\ 10.0\right]  $ & $\left[  5.8,\ 10.0\right]  $ &
$\left[  3.4,\ 10.0\right]  $ & $\left[  3.6,\ 10.0\right]  $ & $\left[
2.0,\ 10.0\right]  $\\\hline\hline
\end{tabular}
\end{center}

\subsection{Tolman VI-like solution}

In this case, the model to be studied is the solution VI of Tolman. We recall
that the equation of state of this model, for large $\rho$, approaches that
for a highly compressed Fermi gas. The density matter is%

\begin{equation}
\rho=\frac{3\,\sigma}{56\,\pi\,r^{2}}%
\end{equation}
Respectively, the mass and pressure are found to be
\begin{equation}
m=\frac{3}{14}\sigma r\qquad\mathrm{and}\qquad P_{r}=-\rho+\frac{\mathcal{C}%
}{2\,\pi r^{3}}\ .
\end{equation}
The constant $C_{1}$ and the boundary conditions lead to
\begin{equation}
{\mathcal{C}}=\frac{3}{28}\,\sigma\,a\qquad\mathrm{and}\qquad\sigma
=\frac{14\,M}{3\,a}%
\end{equation}

The energy conditions yield to

\begin{center}%
\begin{tabular}
[c]{|c|c|c|c|c|c|c|c|c|}\hline\hline
$r\leq10.0$ & $\rho\geq0$ & $P_{r}\geq0$ & $P_{\perp}\geq0$ & $\frac{\partial
P_{r}}{\partial r}<0$ & $\frac{\partial P_{r}}{\partial\rho}\leq1 $ &
$\rho>P_{r}+2\,P_{\perp}$ & $\rho>P_{r}$ & $\rho>P_{\perp}$\\\hline
$r\in$ & $\left(  0,\ 10.0\right]  $ & $\left(  0,\ 10.0\right]  $ & $\left(
0,\ 5.0\right]  $ & $\left(  0,\ 10.0\right]  $ & $\left[  7.5,\ 10.0\right]
$ & $\left[  5.0,\ 10.0\right]  $ & $\left[  5.0,\ 10.0\right]  $ & $\left[
3.0,\ 10.0\right]  $\\\hline\hline
\end{tabular}
\end{center}

\subsection{Stewart Solution}

B.W. Stewart examines several anisotropic, conformally flat, internal
solutions. He presents four examples of mass distributions, we select the
example 2 from that paper \cite{Stewart1982}%

\begin{equation}
\rho=\frac{1}{8\,\pi r^{2}}\frac{\left(  e^{2Kr}-1\right)  \left(
e^{4Kr}+8Kre^{2Kr}-1\right)  }{\left(  e^{2Kr}+1\right)  ^{3}}%
\end{equation}
The expressions for the mass distribution and pressure are
\begin{equation}
m=\frac{r}{2}\left(  \frac{e^{2Kr}-1}{e^{2Kr}+1}\right)  ^{2}\qquad
\mathrm{and}\qquad P_{r}=\rho-\frac{1}{4\pi r^{2}}\left(  \frac{e^{2Kr}
-1}{e^{2Kr}+1}\right)  ^{2}+\frac{\mathcal{C}}{2\,\pi r^{3}}%
\end{equation}

Again, the constant $C_{1}$ is obtained to make $P_{r}(a)=0,$ i.e.
\begin{equation}
{\mathcal{C}}=\frac{a}{4}\frac{e^{2Ka}\left(  e^{3Ka}-8Kae^{2Ka}%
-e^{2Ka}+8Ka-1\right)  +1}{e^{2Ka}\left(  e^{3Ka}+3e^{2Ka}+3\right)  +1}%
\end{equation}
while the constant $K$ is found from the boundary conditions as
\begin{equation}
K=\frac{1}{2a}\ln\left[  \frac{1+\left(  \frac{2M}{a}\right)  ^{\frac{1}{2}}
}{1-\left(  \frac{2M}{a}\right)  ^{\frac{1}{2}}}\right]
\end{equation}

The energy conditions lead

\begin{center}%
\begin{tabular}
[c]{|c|c|c|c|c|c|c|c|c|}\hline\hline
$r\leq10.0$ & $\rho\geq0$ & $P_{r}\geq0$ & $P_{\perp}\geq0$ & $\frac{\partial
P_{r}}{\partial r}<0$ & $\frac{\partial P_{r}}{\partial\rho}\leq1 $ &
$\rho>P_{r}+2\,P_{\perp}$ & $\rho>P_{r}$ & $\rho>P_{\perp}$\\\hline
$r\in$ & $\left(  0,\ 10.0\right]  $ & $\left(  0,\ 10.0\right]  $ & $\left(
0.5,\ 10\right)  $ & $\left(  0,\ 10.0\right]  $ & $\left[  1.9,\ 10.0\right]
$ & $\left[  0.6,\ 10.0\right]  $ & $\left(  0.5,\ 10.0\right)  $ & $\left(
0,\ 10.0\right]  $\\\hline\hline
\end{tabular}
\end{center}

\subsection{Gokhroo and Mehra solution}

Now, the Gokhroo and Mehra\ \cite{GokhrooMehra1994}\ solution is presented.
The mass density is
\begin{equation}
\rho=\sigma\left[  1-K\frac{r^{2}}{a^{2}}\right]
\end{equation}
the mass and the radial pressure, can be written as
\begin{equation}
m={\frac{\sigma\,{r}^{3}}{6}}\,\left[  1\ -\frac{3\,K}{5}\frac{{r}^{2}}
{{a}^{2}}\right]  \qquad\mathrm{and}\qquad P_{r}=\frac{3}{5}\rho-\frac{1}
{2\pi}\left[  \frac{\sigma}{15}-\frac{\mathcal{C}}{r^{3}}\right]
\end{equation}
and the boundary conditions lead to
\begin{equation}
{\mathcal{C}}=\frac{\sigma a^{3}}{60}\left[  9K-5\right]  \qquad\mathrm{and}
\qquad\sigma=\frac{30\,M}{a^{3}\left(  5-3K\right)  }%
\end{equation}

The physical variables are valid in the following range

\begin{center}%
\begin{tabular}
[c]{|c|c|c|c|c|c|c|c|c|}\hline\hline
$r\leq10.0$ & $\rho\geq0$ & $P_{r}\geq0$ & $P_{\perp}\geq0$ & $\frac{\partial
P_{r}}{\partial r}<0$ & $\frac{\partial P_{r}}{\partial\rho}\leq1 $ &
$\rho>P_{r}+2\,P_{\perp}$ & $\rho>P_{r}$ & $\rho>P_{\perp}$\\\hline
$r\in$ & $\left(  0,\ 10.0\right]  $ & $\left(  0,\ 10.0\right]  $ & $\left(
0,\ 8.2\right]  $ & $\left(  0,\ 10.0\right]  $ & $\left[  1.1,10.0\right]  $%
& $\left(  0,\ 10.0\right]  $ & $\left(  0,\ 10.0\right]  $ & $\left(
0,\ 10.0\right]  $\\\hline\hline
\end{tabular}
\end{center}

\subsection{Durgapal Solution}

The paper of M. C. Durgapal \cite{Durgapal1982} report several solutions. The
autor assume that the value of the metric coefficient is given by: $e^{\nu
}=A\left(  1+Cr^{2}\right)  ^{n}$, where $A$ and $C$ are constant and $n$ a
parameter. Durgapal shows that Einstein's equations can be solved explicitly
for different values from n.

\begin{itemize}
\item  For $n=1$

The first solution are identical to Tolman IV solution. The expression for the
density is given by
\begin{equation}
\rho=\frac{C}{8\,\pi}\left[  \frac{1-3K-3Kx}{\left(  1+2\,x\right)  }
+\frac{2\left(  1+Kx\right)  }{\left(  1+2\,x\right)  ^{2}}\right]
\end{equation}
Again, the expressions for mass and radial pressure are obtained from the
equation (\ref{globalst}):
\begin{equation}
m=-\frac{x^{\frac{3}{2}}}{2C^{\frac{1}{2}}}\frac{K\left(  1+x\right)
-1}{\left(  1+2x\right)  }\qquad\mathrm{and}\qquad P_{r}=\rho+\frac{C}{4\,\pi
}\frac{K\left(  1+x\right)  -1}{\left(  1+2x\right)  }+\frac{\mathcal{C}}
{2\,\pi\left(  \frac{x}{C}\right)  ^{\frac{3}{2}}}%
\end{equation}
where
\begin{equation}
x=C\,r;\qquad x_{1}=C\,a^{2}=\frac{M}{a-3M};\qquad\mathrm{and}\qquad
K=\frac{\sqrt{C}\left[  ax_{1}-2M\left(  1+2x_{1}\right)  \right]  }
{x_{1}^{\frac{3}{2}}\left(  1+x_{1}\right)  }%
\end{equation}
The constant $K$ is obtained for boundary conditions. The pressure in the
surface ($P_{r}(a)=0$) leads to that:
\begin{equation}
{\mathcal{C}}=\frac{ax_{1}}{4}\frac{K\left(  1+x_{1}+2x_{1}^{2}\right)
+2x_{1} -1}{1+4x_{1}\left(  1+x_{1}\right)  }%
\end{equation}

The validity range are:%

\begin{tabular}
[c]{|c|c|c|c|c|c|c|c|c|}\hline\hline
$r\leq10.0$ & $\rho\geq0$ & $P_{r}\geq0$ & $P_{\perp}\geq0$ & $\frac{\partial
P_{r}}{\partial r}<0$ & $\frac{\partial P_{r}}{\partial\rho}\leq1 $ &
$\rho>P_{r}+2\,P_{\perp}$ & $\rho>P_{r}$ & $\rho>P_{\perp}$\\\hline
$r\in$ & $\left(  0,\ 10.0\right]  $ & $\left(  0,\ 10.0\right]  $ & $\left(
0,\ 10.0\right]  $ & $\left(  0,\ 10.0\right]  $ & $\left(  0,\ 10.0\right]  $%
& $\left(  0,\ 10.0\right]  $ & $\left(  0,\ 10.0\right]  $ & $\left(
0,\ 10.0\right]  $\\\hline\hline
\end{tabular}

\item  For $n=2$

This solution is identical to that obtained by Kuchowicz, Adler and Adams and
Cohen, according to the author. The density is given by
\begin{equation}
\rho=-\frac{C}{8\,\pi}\frac{K\left(  3+5\,x\right)  }{\left(  1+3\,x\right)
^{\frac{5}{3}}}%
\end{equation}
The mass and radial pressure, are :
\begin{equation}
m=-\frac{1}{2C^{\frac{1}{2}}}\frac{Kx^{\frac{3}{2}}}{\left(  1+3x\right)
^{\frac{2}{3}}}\qquad\mathrm{and}\qquad P_{r}=\rho+\frac{1}{4\,\pi}\frac
{CK}{\left(  1+3x\right)  ^{\frac{2}{3}}}+\frac{\mathcal{C}}{2\,\pi\left(
\frac{x}{C}\right)  ^{\frac{3}{2}}}
\end{equation}
where
\begin{equation}
x=C\,r;\qquad x_{1}=C\,a^{2}=\frac{M}{2a-5M};\qquad\mathrm{and}\qquad
K=-\frac{2\,M}{a\,x_{1}}\left(  1+3\,x_{1}\right)  ^{\frac{2}{3}}
\end{equation}
The constant $K$ is obtained for boundary conditions and ${\mathcal{C}}$ by
($P_{r}(a)=0$)
\begin{equation}
{\mathcal{C}}=\frac{CKa^{3}}{4}\frac{1-x_{1}}{\left(  1+3x_{1}\right)
^{\frac{5}{3}}}%
\end{equation}

The regions where the physical variables are valid%

\begin{tabular}
[c]{|c|c|c|c|c|c|c|c|c|}\hline\hline
$r\leq10.0$ & $\rho\geq0$ & $P_{r}\geq0$ & $P_{\perp}\geq0$ & $\frac{\partial
P_{r}}{\partial r}<0$ & $\frac{\partial P_{r}}{\partial\rho}\leq1 $ &
$\rho>P_{r}+2\,P_{\perp}$ & $\rho>P_{r}$ & $\rho>P_{\perp}$\\\hline
$r\in$ & $\left(  0,\ 10.0\right]  $ & $\left(  0,\ 10.0\right]  $ & $\left(
0,\ 10.0\right]  $ & $\left(  0,\ 10.0\right]  $ & $\left(  0,\ 10.0\right]  $
& $\left(  0,\ 10.0\right]  $ & $\left(  0,\ 10.0\right]  $ & $\left(
0,\ 10.0\right]  $\\\hline\hline
\end{tabular}

\item  For $n=3$

In the third solution studied by Durgapal the density has the following form:
\begin{equation}
\rho=\frac{C}{8\,\pi\left(  1+\,x\right)  ^{2}}\left[  \frac{3}{2}\left(
3+x\right)  -\frac{3K\left(  1+3\,x\right)  }{\left(  1+4\,x\right)
^{\frac{3}{2}}}\right]
\end{equation}

This expression for the density is exactly the same expression for the
densited reported in Durgapal and Bannerji for the case with $K=0$

Again, the mass and the radial pressure are
\begin{equation}
m=-\frac{x^{\frac{3}{2}}}{4C^{\frac{1}{2}}}\frac{2K-3\left(  1+4x\right)
^{\frac{1}{2}}}{\left(  1+x\right)  \left(  1+4x\right)  ^{\frac{1}{2}}
};\qquad\mathrm{and}\qquad P_{r}=\rho+\frac{C}{8\,\pi}\frac{2K-3\left(
1+4x\right)  ^{\frac{1}{2}}}{\left(  1+x\right)  \left(  1+4x\right)
^{\frac{1}{2}}}+\frac{\mathcal{C}}{2\,\pi\left(  \frac{x}{C}\right)
^{\frac{3}{2}}}%
\end{equation}
where
\begin{equation}
x=C\,r;\qquad x_{1}=C\,a^{2}=\frac{M}{3a-7M};\qquad\mathrm{and}\qquad
K=-\frac{\left(  1+4\,x_{1}\right)  ^{\frac{1}{2}}}{2a\,x_{1}}\left[
4M\left(  1+x_{1}\right)  -3\,a\,x_{1}\right]
\end{equation}
Again, te constant $K$ is obtained for boundary conditions and ${\mathcal{C}}$
by: ($P_{r}(a)=0$)
\begin{equation}
{\mathcal{C}}=\frac{x_{1}^{\frac{3}{2}}}{8C^{\frac{1}{2}}}\frac{2K\left(
1-x_{1}-8x_{1}^{2}\right)  -3\left(  1+4x_{1}\right)  ^{\frac{1}{2}}\left(
1+3x_{1}-4x_{1}^{2}\right)  }{\left(  1+4x_{1}\right)  ^{\frac{3}{2}}\left(
1+2x_{1}+x_{1}^{2}\right)  }%
\end{equation}

The validity ranks are%

\begin{tabular}
[c]{|c|c|c|c|c|c|c|c|c|}\hline\hline
$r\leq10.0$ & $\rho\geq0$ & $P_{r}\geq0$ & $P_{\perp}\geq0$ & $\frac{\partial
P_{r}}{\partial r}<0$ & $\frac{\partial P_{r}}{\partial\rho}\leq1 $ &
$\rho>P_{r}+2\,P_{\perp}$ & $\rho>P_{r}$ & $\rho>P_{\perp}$\\\hline
$r\in$ & $\left(  0,\ 10.0\right]  $ & $\left(  0,\ 10.0\right]  $ & $\left(
0,\ 10.0\right]  $ & $\left(  0,\ 10.0\right]  $ & $\left(  0,\ 10.0\right]  $
& $\left(  0,\ 10.0\right]  $ & $\left(  0,\ 10.0\right]  $ & $\left(
0,\ 10.0\right]  $\\\hline\hline
\end{tabular}
\end{itemize}

\end{document}